# Molecular Brightness analysis of GPCR oligomerization in the presence of spatial heterogeneity

Paolo Annibale*, Martin J Lohse
Max Delbrück Center for Molecular Medicine in the Helmholtz Association, Berlin

paolo.annibale@mdc-berlin.de

Measuring the oligomerization of plasma membrane proteins is rife with biophysical and biomedical implications. This is particularly true for GPCRs, a large family of proteins representing the targets of over one third of all FDA approved medications. Over the last thirty years, fluorescence has been the leading approach to address this problem. However, in spite of a large number of studies and approaches, for most GPCRs the results have remained contentious, possibly due to the large spectrum of specific methods employed[1,2], besides biological diversity.

Stoneman et al. recently presented a novel approach to extract the oligomerization state of fluorescently labeled membrane proteins from microscopy images, including the class B GPCR secretin receptor[3]. Their approach is ultimately based on the technique known as Spatial Intensity Distribution Analysis (SpIDA), which belongs to the family of molecular brightness (MB) methods[4-7]. As opposed to single molecule studies, where the oligomerization state of a fluorescent species can be extracted from the intensities of the individual molecular 'spots' observed with the camera, in MB a statistical average is extracted, based on the width of the histogram of the measured intensity counts. This can be performed in a spatial (as Stoneman *et al.* did) or temporal fashion, as originally done using fluorescence correlation spectroscopy (FCS)[6]. In the end, the MB $\varepsilon$, a property of the individual fluorescent species relating to its *average* oligomerization state, is extracted from the histogram: putting it simple, a dimer is twice as bright as a monomer, since it carries two fluorophores.

The key advance offered by Stoneman et al. is that they divide the image into many small regions (ROIs of the order of about a $\mu m^2$), and each ROI is then analyzed independently and provides one single MB value reflecting the average oligomerization state in that ROI. In order to recapitulate this approach, we use here the original dataset provided by Stoneman *et al*. and determine the MB in 36 ROIs on the membrane of a cell expressing membrane-targeted monomeric EGFP, as indicated in region A of **Figure 1a**. Each ROI is color coded according to the average oligomerization state along the line of analysis by Stoneman *et al (*see **Supplementary Information**). All MBs from the ROIs are then combined, yielding a histogram of brightness values from multiple, distinct ROIs, as schematically depicted in the inset of **Figure 1a**. Such a histogram is then fitted with multiple Gaussian peaks in order to reconstruct different oligomeric populations (monomers, dimers...), very much alike to what is done in single particle tracking approaches[2,8].

In our opinion, this approach is based upon and opens up a very interesting issue concerning biologically relevant length-scales at the plasma membrane: if multiple peaks in a brightness histogram derived from such ROI analysis arise, then the distinct levels of oligomerization (monomers, dimers…) must be different from one ROI to the other, i.e. be segregated in space (**SI Figure 1**). In this view, important signaling molecules such as GPCRs may form monomers at one position of the plasma membrane of a cell but oligomers at another position. Such regional differences would have important implications, and we therefore set out to investigate this aspect further.

Let us observe a representative, randomly chosen (**region B in Figure 1a**), zoom-in of the basolateral membrane of Stoneman's HEK293 cell expressing the membrane-bound mEGFP. As we reach the length-scale (~1.25 μm x 1.25 μm) of the ROIs used in their work (**Figure 1b-c**), we observe that the plasma membrane appearance remains far from homogeneous: intensity *hot-spots* of varying size and intensity occur in almost every ROI. When we measure the MB (one value per ROI as in Stoneman's approach) in **Figure 1e**, a significant heterogeneity is seen. Notably, among two of the four ROIs there is 1.5-fold difference in MB. Since each ROI contains hundreds of fluorescently labeled molecules, this would imply that in one ROI these molecules are predominantly in a *monomeric* state, whereas in another ROI, only one micron apart, there are now hundreds of molecules which are in a non-monomeric state, i.e. dimers or higher order oligomers.

The presence of brightness hotspots can be also seen by a temporal brightness acquisition: **SI Figure 2a-c** shows a representative frame of a movie of the basolateral membrane of a HEK293 cell expressing the $β_1$-AR, a prototypical, largely monomeric class A GPCR[8]. From confocal time sequences, temporal brightness values for each pixel can be extracted, as illustrated in **SI Figure 2d**. Here, we observe a sizable hot-spot which is more than 8 times brighter than the baseline values.

The question now is whether these brightness hot-spots result from spatially heterogeneous oligomerization of the membrane proteins or from other factors. The first possibility entails a significant rethinking of the basic hypothesis of the Fluid Mosaic Model[9], namely that plasma membrane proteins can diffuse freely and that the law of mass action applies homogeneously. Significant evidence already points to the very heterogeneous nature of the membrane, and Stoneman's work would provide unprecedented evidence in this direction, suggesting that in some spots at the cell surface a protein may be monomeric while in other spots it may be di- or oligomeric.

On the other hand, one may assume that, depending on the length-scale investigated, the cell membrane presents a significant degree of *morphological* and *functional* heterogeneity. Notably, this might be inferred from the increase of measured MB as a function of the size of the area investigated (**Figure 1f**), reflecting an increase in MB due to the growing spatial heterogeneity. If we look at confocal snapshots of another prototypical GPCR, namely the $β_2$-AR, which undergoes constitutive internalization, we can observe large endosomes moving near the cell surface (arrowheads in **Figure 1g-h**), endocytic pits (**Figure 1i**) or microscopic membrane leaflets and filopodia under the cell; in certain instances also receptors within subplasmalemmal portions of the endoplasmic reticulum can contribute to the contrast (**Figure 1l**).

Ideally, an alternative approach to analyze such data would be to carefully avoid such hot-spots and to focus on regions of the basolateral membrane that are as homogeneous as possible. In such homogeneous regions, photon count fluctuations are more likely to arise from the stochastic fluctuations of protomers diffusing and potentially oligomerizing. Given the extreme heterogeneity in sizes, intensity values and dynamics of these hot-spots (also **SI Figure 3** for the $β_2$-AR), this may be easier said than done, leaving it a significant challenge in the field. In our example, after manual selection of a ROI (as illustrated in **Figure 1m**), it appears that a MB value closer to the monomer is recovered, suggesting that the large hot-spots, now excluded from the analysis, do not directly relate to bona-fide receptor oligomerization, but rather to processes on a larger spatial scale. Accordingly, analysis of simulated datasets containing homogeneous mixture of monomers and dimers, also yields a more homogeneous distribution of MB values (**Supplementary Note 3**).

While SpIDA, which represents a static snapshot of the plasma membrane, cannot discriminate a large, immobile background spot from a large oligomer, temporal brightness measures in each pixel fluctuations over time around the average, and hence provides a correction for these potential artifacts. On the other hand, it is more prone to overestimating MB in the presence of highly dynamic background features.

In our view, it appears that by *combining* the two methods and interpretations together with a careful selection of the area to be investigated, one may perform even more robust oligomerization studies[10]. In the light of these results, the high-throughput MB approach of Stoneman et al. may provide interesting results where a spatially polarized or asymmetric distribution of oligomerization states may be present (e.g. neurons). While a fit-it-all solution to extract oligomerization at the plasma membrane appears to be still missing, in our opinion the combination of multiple methods and a careful inspection of the area under investigation - in order to exclude phenomena which cannot be strictly attributed to molecular oligomerization -, may help propelling the field further.

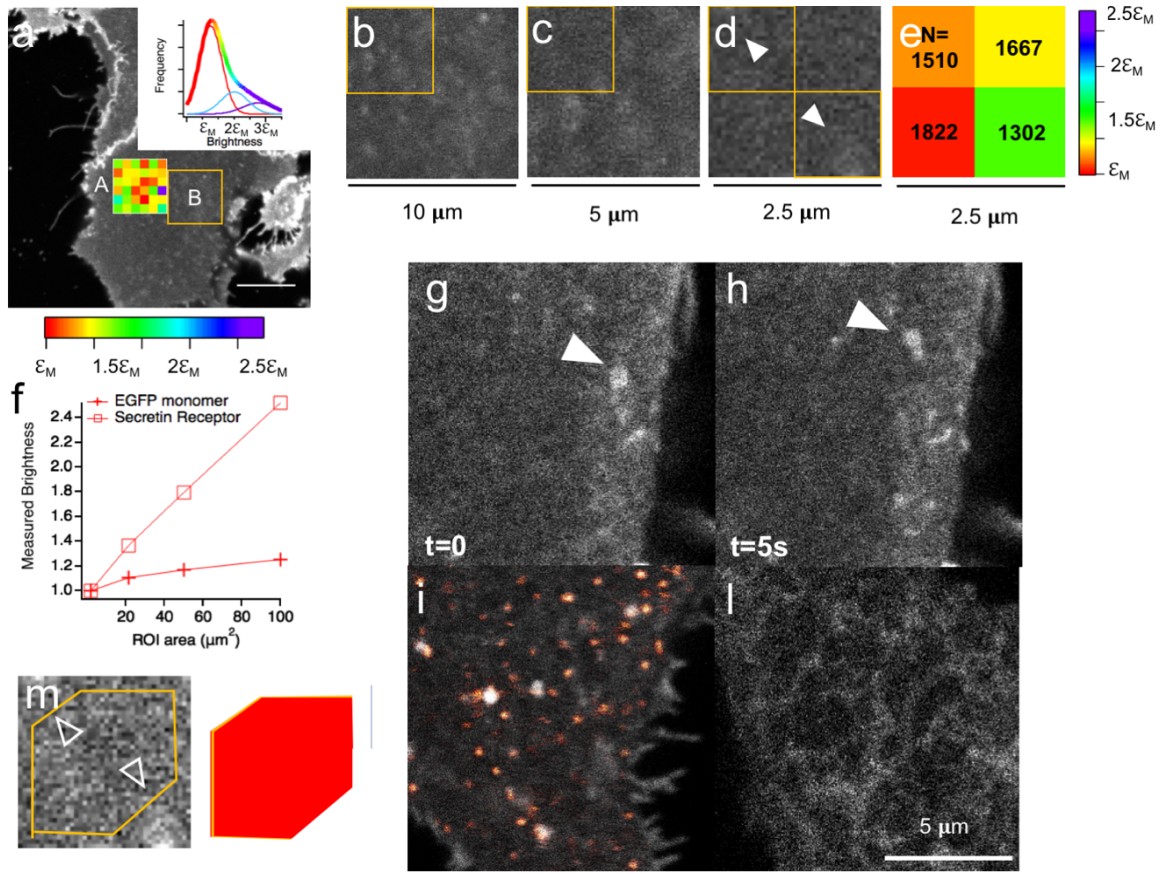

*Figure 1* **a)** Portion of the plasma membrane of a Flp-In T-REx HEK293 cell expressing PM-1-mEGFP (adapted from Fig. 1 of Stoneman *et al.*[3]). Overlaid (A) a subset of the ROIs used to construct the histogram of the MB (inset). We measured the MB in each ROI (see Supplementary Information) and color coded it to reflect the predominant oligomerization state (see scale at right). **b-d)** Region B in panel a, for increasing magnifications. White arrows indicate intensity hot-spots. **e)** Color coded MB in the four ROIs highlighted in d; overlaid is the number of molecular complexes found in each ROI. **f)** The average MB (fold-change) for monomeric GFP and secretin receptor (from the data of Stoneman *et al.*) increases as the size of the ROI is increased from ≈1.25 to 100 $\mu m^2$ (n=3). **g-i)** Confocal micrographs of HEK293 cells expressing $\beta_2$-AR C-terminally fused to EYFP. **g)** Arrow indicates a moving endosome, as highlighted by a subsequent micrograph **h)** taken after 5s. **i)** $\beta_2$-AR undergoing internalization (5 minutes after addition of the agonist isoproterenol), overlaid to clathrin light chains stained with mCherry. **l)** Outline of the subplasmalemmal endoplasmic reticulum, illustrating partial retention of the $\beta_2$-AR. **m)** (left) Panel d with enhanced contrast. (right) MB recovered ($\epsilon=1.16\epsilon_M$, N=1046) by an area selection which excludes intensity hot-spots, as indicated by the arrows. The color code is the same as in e.

1   Ferre, S. *et al.* G Protein-Coupled Receptor Oligomerization Revisited: Functional and Pharmacological Perspectives. *Pharmacol Rev* **66**, 413-434, doi:10.1124/pr.113.008052 (2014).

2       Scarselli, M. *et al.* Revealing G-protein-coupled receptor oligomerization at the single-molecule level through a nanoscopic lens: methods, dynamics and biological function. *Febs J* **283**, 1197-1217, doi:10.1111/febs.13577 (2016).

3       Stoneman, M. R. *et al.* A general method to quantify ligand-driven oligomerization from fluorescence-based images. *Nat Methods* **16**, 493-496, doi:10.1038/s41592-019-0408-9 (2019).

4       Godin, A. G. *et al.* Revealing protein oligomerization and densities in situ using spatial intensity distribution analysis. *Proceedings of the National Academy of Sciences of the United States of America* **108**, 7010-7015, doi:10.1073/pnas.1018658108 (2011).

5       Qian, H. & Elson, E. L. Distribution of Molecular Aggregation by Analysis of Fluctuation Moments. *P Natl Acad Sci USA* **87**, 5479-5483, doi:DOI 10.1073/pnas.87.14.5479 (1990).

6       Chen, Y., Muller, J. D., So, P. T. C. & Gratton, E. The photon counting histogram in fluorescence fluctuation spectroscopy. *Biophys J* **77**, 553-567, doi:Doi 10.1016/S0006-3495(99)76912-2 (1999).

7       Digman, M. A., Dalal, R., Horwitz, A. F. & Gratton, E. Mapping the number of molecules and brightness in the laser scanning microscope. *Biophys J* **94**, 2320-2332, doi:10.1529/biophysj.107.114645 (2008).

8       Calebiro, D. *et al.* Single-molecule analysis of fluorescently labeled G-protein-coupled receptors reveals complexes with distinct dynamics and organization. *Proceedings of the National Academy of Sciences of the United States of America* **110**, 743-748, doi:10.1073/pnas.1205798110 (2013).

9       Singer, S. J. & Nicolson, G. L. The Fluid Mosaic Model of the Structure of Cell Membranes. *Science* **175**, 720-+, doi:DOI 10.1126/science.175.4023.720 (1972).

10      Serfling, R. *et al.* Quantitative Single-Residue Bioorthogonal Labeling of G Protein-Coupled Receptors in Live Cells. *ACS Chem Biol*, doi:10.1021/acschembio.8b01115 (2019).

# Supplementary Information:
Molecular Brightness analysis of GPCR oligomerization in the presence of spatial heterogeneity
Paolo Annibale, Martin Lohse

## Analysis of the data reported in Stoneman et al.

<u>Interpretation of the Molecular Brightness Histograms as in Stoneman et al.</u>

Stoneman et al.[1] have developed a method whereby they divide their image into micrometer-sized segments/Regions of Interest (ROI), and for each of these regions they evaluate the intensity distributions of the pixels in order to extract the molecular brightness (MB). The brightness values for all the segments/ROIs are finally collected and put into a histogram. The key assumption is that this broad MB histogram originates from the sum of peaks, each representing one of the oligomeric components (monomer, dimer, trimer etc) present (and possibly mixed) at the cell membrane.

We recapitulated this process graphically in **Figure 1a** and its inset. We would like to point out that our line of reasoning relies entirely on an examination of the primary data presented in Stoneman et al.[1]

Considering that the measured MB comes from a combination of the brightness of each species present (monomer, dimer…) weighted by its concentration (eq. S33 in Stoneman et al.[1]), this implies that a segment/ROI corresponding to a bin at the center of the dimeric peak in the MB histogram, is populated by a majority of dimers. This is schematically illustrated in **SI Figure 1**.

To make an example, in a sample containing a monomer/dimer equilibrium (and if one were to ignore any other source of noise or error) a MB value exactly twice that of the monomer can occur in a segment/ROI only if all (or the great majority) of the molecules in the segment/ROI are dimers, or a very precise combination of monomers and higher order oligomers, e.g. exactly 80% monomers and 20% tetramers etc. This is illustrated in **SI Figure 4**, based on equation *S33* of Stoneman et al.[1]

<u>Relationship between MB and intensity histograms in each individual segment/ROI as in Stoneman et al.</u>

We recapitulate in **SI Figure 5** the steps leading to the generation of the MB value for a given segment/ROI starting from the raw image data. As an example, we use the four ROIs originally displayed in **Figure 1e**, originating from the same cell reported in *Figure 1a* of Stoneman et al., namely a HEK293 cell expressing GPI-anchored mEGFP (used as a monomeric control). The ROIs have approximately the same size as those reported by Stoneman et al., (side≃1.25 µm).

First, the histogram of the pixel intensity values within each ROI is calculated (**SI Figure 5a,b**). Then, a model is fit to the histogram, containing the MB $\varepsilon$ as a fit parameter. In most cases (including Stoneman et al.) and in particular when a large number of molecules is present in the ROI, a Gaussian curve is fit to such histogram. When a Gaussian curve is used, the MB $\varepsilon$ directly relates to the variance of the Gaussian curve, $\sigma^2$ (eqs. S1-S5 in Stoneman et al.[1], and also[2,3]). The number of molecules N present in the ROI is obtained by dividing the average intensity <I> within the ROI by $\varepsilon$ (**SI Figure 5c**).

One single MB value $\varepsilon$ and number of molecules N are then associated to each segment/ROI, as illustrated in **SI Figure 5d**.

## Materials and Methods

Confocal acquisitions

Confocal micrographs were acquired using either a Leica SP5 or a Leica SP8 confocal microscope using HyD detectors in photon counting mode. Cells were imaged with a 40x / 1.25 numerical aperture oil immersion objective.

Measurement of Temporal Brightness

Temporal Brightness Measurements were performed as previously described[4]. Briefly, the imaging mode was XYT (2D and time) and 100 frames were taken with a scanner speed of 400 Hz using the following parameters: pinhole-size: 67.93/ zoom-factor: 30.3. Resolution 256x256 pixels. EYFP-tagged constructs were imaged using a 514 nm line of an Arg-Ion laser at a power of 2.5%. Data were analyzed using a custom-written IgorPro (Wavemetrics) routine as described previously[5]. The brightness values were calculated based on the average of the brightness values from each pixel within the region of interest[6].

Measure of Spatial Brightness

Image analysis was performed applying existing and publicly available software for Spatial Intensity Distribution Analysis (SpIDA)[3]. The decision to use an algorithm previously published to extract MB values from a ROI of a confocal micrograph is based on the fact that the ultimate foundations of Stoneman et al. method (extracting MB from the histogram of the intensity counts of a ROI/segment) and that of the previously published SpIDA approach (extracting MB from the histogram of the intensity counts of a ROI/segment) are the same.
The absolute MB values depend on the specific algorithm used, as well as the parameters used as inputs in the algorithm. When an analog detector is used, the properties and settings of the detector, namely dark counts, S-factor, gain, variance of the dark counts and linear range, shall be further included in the analysis. However, in this case and as discussed more in detail in **Supplementary Note 2**, the choice of different algorithms to extract the MB affects only the absolute MB values recovered, but overall conserves relative MB values across different ROIs/segments. This reflects the fact that, no matter what fitting function or algorithm is used to extract the MB, ROIs/segments displaying a higher variance of the pixel intensities (for comparable averages, and same detector settings), will yield a higher MB value.

## Supplementary Notes

Supplementary Note 1: Comparison of different approaches to extract spatial MB values

**SI Table 1** displays the MB values extracted for the four exemplary ROIs presented in **Figure 1e** and **SI Figure 5**, using three different approaches.

1. First, we performed a single Gaussian fit to the pixel intensity histograms (**SI Figure 5b**) and reported the mean and standard deviation (stdev).
2. Then, we used the SpIDA algorithm assuming an ideal photon counting detector (the full characterization of the detector performance, e.g. dark counts level or linearity, was unfortunately not available in Stoneman et al., see **Supplementary Note 2**).
3. Further, we used the SpIDA algorithm taking into account the analog detection: offset=50 counts (estimated), S=37, $\sigma_0$= 0. The linearity range of the detector was not provided.

As an indicator of how heterogeneous are the MB values recovered across the four ROIs by each method, the ratio of the MB of each ROI to the lowest value is reported. **SI Table 1** illustrates that, notwithstanding the changes in absolute MB value, the ratios between the ROIs are conserved for different approaches. ROI3 (bottom left) displays consistently the lowest variance/brightness values, whereas ROI4 (bottom right) - which contains a visible intensity 'blob' - yields the largest value. The direct measurement of the standard deviation (stdev) of the intensity histograms yields the smallest ratios, whereas the use of the SpIDA algorithm adjusted for the properties of the analog detector gives the largest maximum to minimum MB fluctuation (1.7) between the four ROIs.

Supplementary Note 2: effect of analog detector characteristics on the recovered spatial MB

It is generally true that detector noise affects the estimation of the MB: for this reason, it is generally advisable to work with photon counting detectors, which represent the state of the art and remove much of the extra steps related to detector calibration[4-6]. When forced to use an analog detector, the distribution of *intensity* counts becomes 1. shifted to the right (because of detector dark counts), 2. 'fatter' (because of the variance of detector noise adds to that due to true molecular fluctuations) and 3. stretched (because of amplification gain) with respect to the photon counts distribution. As discussed by Godin et al.[3], in SpIDA this results in convolving a detector gaussian with the histogram of the true photon counts. This makes the *apparent brightness* measured in analog mode larger than the true MB. However, here we are only interested in relative changes of brightness between neighboring regions of the same basolateral membrane. In this context, point 1 does not matter, because the shift is always the same, and points 2 and 3 affects each ROI equally as the brightness effects due to amplification and readout from the detector are common. Therefore, relative changes in brightness arise from fluctuations in signal due to diffusion of the molecules.

More quantitatively, let us take eq. S2 from Stoneman et al.[1] This equation is derived from eq. 4 of Unruh et al.[7]. Notably, the original equation was calculated for an EMCCD-based system where the fluctuations in intensity coming from an individual pixel over time were analyzed (i.e fluctuations over time, not space). In Stoneman et al.[1], remarkably, the equation is used for the case of fluctuations across neighboring pixels within their segments/ROIs/. Assuming that the use of the equation is appropriate in this context, and taking note that - for the data reported in *Figure 1* of Stoneman et al. - $\sigma_0$= 0 we are left with the following relationship between variance of the intensity counts and brightness:

$$\sigma^2 = S(\langle I \rangle - offset) + (\langle I \rangle - offset)G\gamma\epsilon$$

Now, looking at the image pixel intensities outside the cells shown in Stoneman et al.[1], offset values can be safely estimated as being of the order of a few percent of the average intensity <I>.

Therefore, the equation can be re-written as:
$$\frac{\sigma^2}{\langle I \rangle} \approx S + G\gamma\epsilon$$
Considering that the SpIDA algorithm (for S=1) outputs the so-called apparent brightness B=$\frac{\sigma^2}{\langle I \rangle}$, the brightness scales with the *true* MB + an offset.

Supplementary Note 3: MB histograms reconstructed from simulated confocal microscopy images containing either monomeric species or 1:1 monomer:dimer mixture.

We generated two sets of simulated confocal microscopy images, yielding each 1000 ROIs/segments of 1.6 µm in size (=32 pixels). The average number of molecules was set at N=400/ROI, yielding a density of approx. 160 protomers/µm$^2$, in the lower range of what is observed in the datasets of Stoneman et al. [1]. In both sets the distribution of the molecules is random. The simulated confocal acquisition has a pixel dwell time of 12.5 µs, and a line retrace time of 5 ms. The excitation PSF has a waist $w_0$=250 nm and $w_z$=750 nm. The molecules diffuse in 2D (=plasma membrane) with D=0.1 µm$^2$/s, in agreement with what is normally reported for membrane receptors.
Each molecule is represented by its Point Spread Function, and photon shot noise is added by using for each pixel an intensity value originating from a Poisson Distribution with mean equal to the original pixel intensity value. Detection is Photon Counting, and no other sources of background or noise affecting the intensity values are included (e.g. quenching, dipole orientation etc.), making these datasets a basis to test an algorithm.

The first set (**SI Figure 6a**) contains only monomers (400 per ROI), with a defined brightness of 0.5 10$^6$ counts/s. In the second set the protomers are equally divided between monomers and dimers, e.g. there are 200 monomers and 100 dimers (having double brightness) (**SI Figure 6b**).

When all the 1000 ROIs/segments of this ideal dataset are analyzed, and the MB of each of them calculated, we are in a position to reconstruct an ideal brightness histogram.
In Stoneman et al.[1] approach, where the MB histogram should represent a spectrum of the brightness/oligomerization states present in the sample, such a histogram should display a unique monomeric peak for the monomer set, and instead reveal the presence of a shoulder, i.e. the dimer peak, for the monomer-dimer set, as in Stoneman's *Figure 1d or SI Figure 1d*. **SI Figure 6c** displays the overlay of the histograms of the raw variances of the pixel intensities calculated for each of the 1000 ROI/segment, for both sets. Although, as expected, the dimeric sample displays an average larger variance, the two curves are functionally identical. There is no evidence of a dimeric shoulder in the monomer/dimer set as opposed to the monomer set.

The same can be observed for the histograms of the MB values recovered using SpIDA in **SI Figure 6d**. We shall note that a brightness of 6 photons/dwell time is recovered for the monomeric sample and about 9 photons/dwell time for the dimeric sample, in agreement with the MB value defined in the simulation for the monomer (500,000 counts/s). The histogram of the monomer/dimer does not display, compared to the monomer set, a peak, shoulder nor even an inflection reflecting the dimeric component. This suggests that when spatial heterogeneities are absent 'by default' and the monomers/dimers homogeneously dispersed and mixed, it is not possible to appreciate/deconvolve oligomeric peak(s) in the histogram of the molecular brightness, but just a shift to the right and a broadening of such histograms.

# Supplementary Tables and Figures

| Annibale et Lohse Figure 1 e | just gauss fit | | ratio to minimum | SpIDA no PMT | | ratio to minimum | SpIDA PMT noise | | ratio to minimum |
|---|---|---|---|---|---|---|---|---|---|
| | mean | stdev | | molecules/pixel | brightness | | molecules/pixel | brightness | |
| bottom left (3) | 1100 ± 10 | 240 ± 20 | 1.00 | 35 | 31 | 1 | 54 | 20 | 1 |
| bottom right (4) | 1140 ± 10 | 330 ± 40 | 1.38 | 26 | 45 | 1.45 | 33 | 34 | 1.7 |
| top left (1) | 1130 ± 10 | 300 ± 20 | 1.25 | 29 | 38 | 1.23 | 42 | 27 | 1.35 |
| top right (2) | 1140 ± 10 | 250 ± 30 | 1.04 | 32 | 36 | 1.16 | 48 | 23 | 1.15 |

SI Table 1

Recovered MB values in four representative ROIs using four methods.

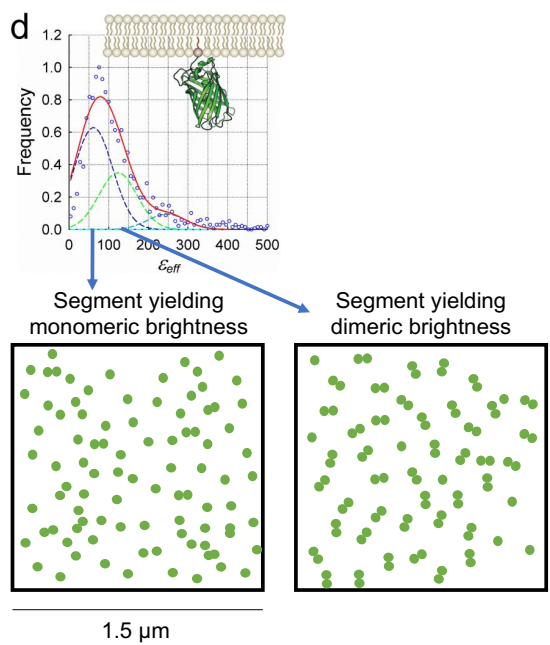

SI Figure 1

Illustration of the data reduction process in two-dimensional Fluorescence Fluctuation Spectrometry (2D-FFS) using two-photon excitation.

**a,** Typical fluorescence image (out of 42 images comprising 146 cells) obtained from two-photon excitation of Flp-In™ T-REx™ cells expressing fluorescently labeled plasma membrane targeted mEGFP construct (PM-1-mEGFP), and an overlaid polygon (P131) indicating a region of interest (ROI) which comprises a patch of the basolateral membrane of a cell. **b,** Software-generated image segmentation of the ROI in (a) using the Moving Square method (see Methods and Supplementary Note 3). **c,** A fluorescence intensity histogram (green circles) of an image segment selected at random, alongside the Gaussian curve (solid red line) used to fit the experimental histogram by adjusting the mean and width of the Gaussian. The intensity binning was set to 25 intensity counts (in arbitrary units). **d-e,** Normalized frequency distribution obtained from the **(d)** PM-1-mEGFP expressing cells (2,803 total segments) was simultaneously fit (solid red line) along with a distribution **(e)** constructed similarly from measurements of cells expressing dimeric, tandem linked mEGFP constructs (2,832 total segments) using a sum of Gaussian functions in order to find brightness of single protomers of mEGFP, $\varepsilon_{eff}^{prot} = 61.4$, when measured using the two-photon optical micro-spectroscope.

MB Histogram for mEGFP imaged in 2-photon excitation (adapted from SI Figure1 of Stoneman et al.[1]), together with a schematic rendering of the individual molecules oligomeric arrangement in segments/ROIs yielding the bins at the center of the MB Histogram monomer (blue dashed) and dimer peak (green dashed) peaks.

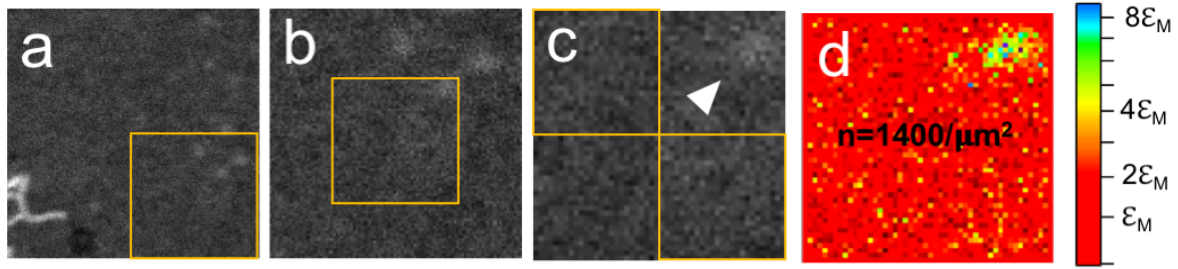

SI Figure 2

Zoom-ins of one frame of the movie of the basolateral membrane of a HEK293 cell expressing the β1-AR c-terminally tagged with EYFP. a) 10 μm square region of the basolateral membrane, b) 5 μm zoom in and b) 2.5 μm zoom in with outlined ROIs comparable in size to those used in Stoneman et al. d) Pixel by Pixel MB extracted, with superposed number of molecules in the ROI.

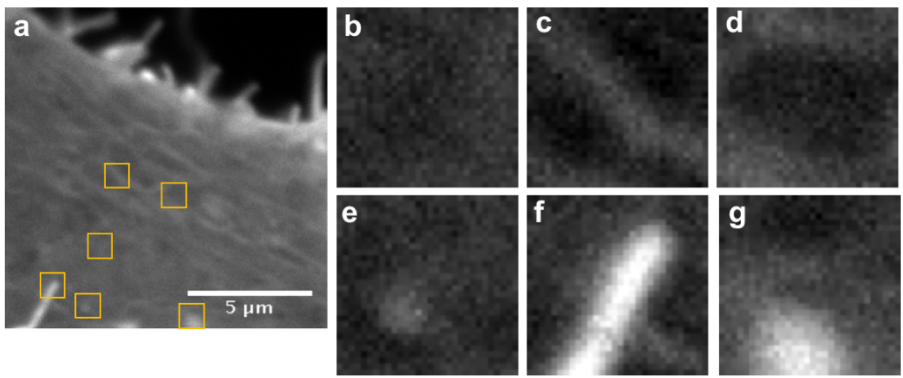

SI Figure 3

Examples of plasma membrane heterogeneities in the distribution of the prototypical GPCR β2-Adrenergic Receptor. a) Confocal image of the basolateral membrane of a HEK293 cell expressing β2-AR-EYFP. Zoom-ins showing respectively a b) homogeneous region of the plasma membrane c) a tubular structure, possibly subplasmalemmal Endoplasmic Reticulum, c) a gap in the ER network d) a small endosome or clathrin coated pit e) the tip of a filopodium twisted under the plasma membrane f) a large endosome. Panels (b-g) are adjusted to the same contrast level.

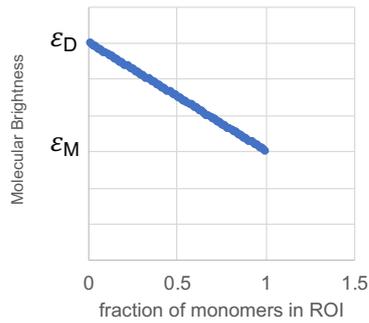

$$\varepsilon_{eff}^{proto}\left(2-\frac{[m]}{c}\right)$$

SI Figure 4

Theoretical dependence of the MB upon the fraction of monomers [m]/c in the ROI. Here $\varepsilon_M=\varepsilon_{eff}^{proto}$ in the notation used in Stoneman et al.[1]

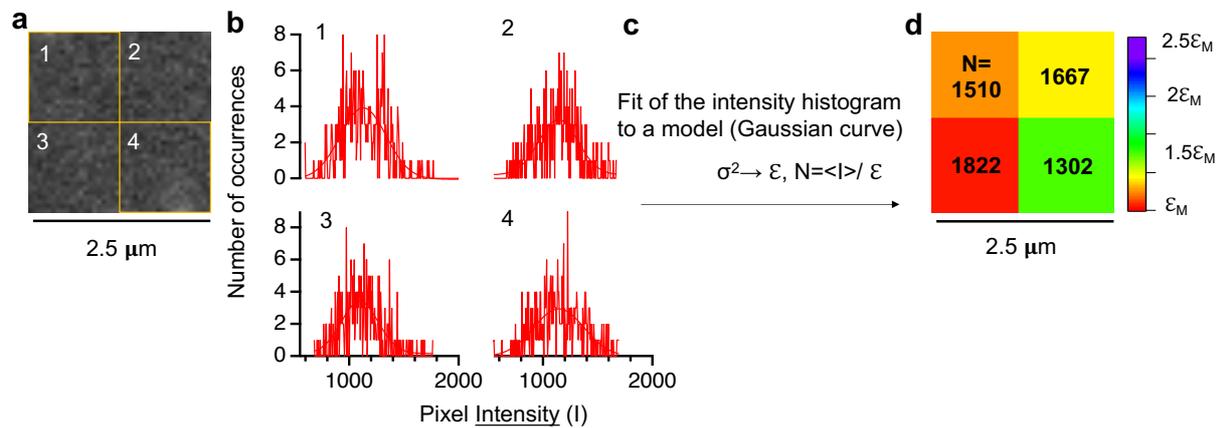

SI Figure 5

Extracting the MB from individual segments/ROIs. a) The four ROIs from Figure 1e are displayed and numbered. b) For each ROI the intensity histogram is calculated and c) Fit to a model in order to extract a MB for the whole segment/ROI. If a Gaussian model is chosen to fit the Intensity Histogram, then the MB directly relates to the variance of the fitting Gaussian curve. d) The MB of each ROI is calculated and shown according to the color-code. The number of molecules N in each ROI is calculated by dividing the average intensity of the ROI by the MB.

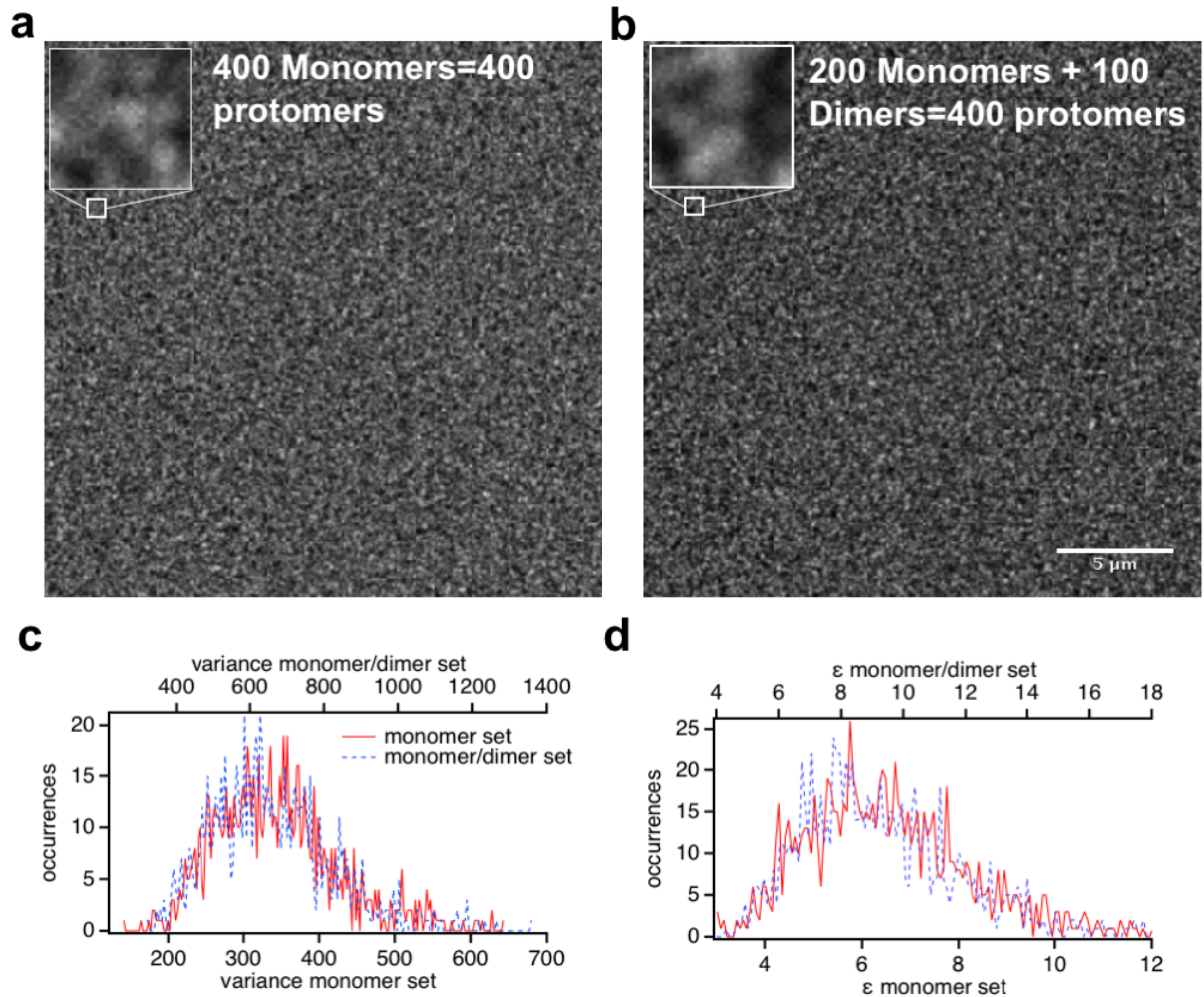

SI Figure 6

MB extracted from simulated monomeric and monomeric/dimeric datasets. a) Montage (30x30) of 900 out of 1000 ROIs/segments containing 160 protomers/µm$^2$, all monomeric, with dashed lines highlighting the size/position of four representative individual ROIs and b) 160 protomers/µm$^2$ equally divided between monomers and dimers. c) Overlap of the Histogram of the variances of pixel intensities measured on each ROI/segment for both monomer (red solid) and monomer/dimer (blue dashed) sets. d) Overlap of the Histogram of the Molecular Brightness (recovered from SpIDA) of each ROI/segment for both monomer (red solid) and monomer/dimer (blue dashed) sets.

# References

1. Stoneman, M. R. *et al.* A general method to quantify ligand-driven oligomerization from fluorescence-based images. *Nat Methods* **16**, 493-496, doi:10.1038/s41592-019-0408-9 (2019).
2. Chen, Y., Muller, J. D., So, P. T. C. & Gratton, E. The photon counting histogram in fluorescence fluctuation spectroscopy. *Biophys J* **77**, 553-567, doi:Doi 10.1016/S0006-3495(99)76912-2 (1999).
3. Godin, A. G. *et al.* Revealing protein oligomerization and densities in situ using spatial intensity distribution analysis. *Proceedings of the National Academy of Sciences of the United States of America* **108**, 7010-7015, doi:10.1073/pnas.1018658108 (2011).
4. Serfling, R. *et al.* Quantitative Single-Residue Bioorthogonal Labeling of G Protein-Coupled Receptors in Live Cells. *ACS Chem Biol*, doi:10.1021/acschembio.8b01115 (2019).
5. Serebryannyy, L. A. *et al.* Persistent nuclear actin filaments inhibit transcription by RNA polymerase II. *J Cell Sci* **129**, 3412-3425, doi:10.1242/jcs.195867 (2016).
6. Isbilir, A. *et al.* Visualization of class A GPCR oligomerization by image-based fluorescence fluctuation spectroscopy. 240903, doi:10.1101/240903 %J bioRxiv (2017).
7. Unruh, J. R. & Gratton, E. Analysis of molecular concentration and brightness from fluorescence fluctuation data with an electron multiplied CCD camera. *Biophysical journal* **95**, 5385-5398, doi:10.1529/biophysj.108.130310 (2008).